\overfullrule=0mm 
\magnification=1200 
\hsize=6.0 truein 
\baselineskip=12pt


\bigskip
\leftline{\bf 1. Introduction}
\medskip

Recently, we adopted the dynamic theory of uniaxial liquid crystals 
(Ericksen-Leslie's theory) to a two-dimensional thin film so as to
make the theory applicable for the study of hydrodynamic behaviour 
of lipid bilayer membranes [1]. 
The membrane was considered to be isolated from spatial surroundings, 
which is evidently impossible to realize in experiments. In aqueous 
solutions lipid membranes tend to close and form vesicles, due to
the spontaneous curvature, maintaining a difference of pressure 
between the interior and exterior liquids. Under the action 
of chemical or electrochemical fields, solutes of the adjacent solutions 
may migrate across the membrane, inducing oscillations of the membrane
potential [2-4], whereas under the action of physical fields, shape or
structure of the membrane may change and may keep varying for a long 
duration [5-8]. Such mass or momentum exchange between the membrane 
and its surroundings is, indeed, essential for the complex dynamic 
processes. The purpose of this paper is to connect in theory the film 
of liquid crystals with the bulk phases of normal fluids.

The theory will be extended to the membrane-solution system in the 
following way: one omits the thickness of the membranes, as in 
the foregoing paper [1], and thus is able to regard the two bulk 
phases as if they were in direct contact with each other. It 
follows that the ``dividing surface'' of the two liquids coincides 
with the ``middle surface'' of the membrane. In any irreversible 
process, on the one hand, kinematic discontinuity 
occurs at the ``dividing surface'' of the bulks, as it is expressed by 
the condition of compatibility (Kotchine's theorem [9]); on the 
other hand, instantaneous balance of mass, momentum and energy is 
required at the ``middle surface'' of the membrane, as it is presented 
in the context of isolated membranes [1]. It stands then to reason
that a superposition of these individual laws gives the description 
of the desired dynamical system. It is done in the second section. 
A group of dynamic equations for an open (and closed) membrane 
system is thus established.

Here we explicitly note that the lipid crystal membrane is a
two-dimensional fluid, which responds to lateral forces with the
surface tension (or compressibility) and to transverse forces with
the elasticity, having no elastic response to shear stretch. This is 
in contrast to the solid film model for red cell membranes suggested 
by Evans [10]. In the gel state, the lateral fluidity vanishes and 
the shape of the membrane varies in the bending mode (if no 
molecule tilting). 

Recent experiments give support to the adoption of liquid crystal 
theory to the study of biomembranes. It was found that 
membrane deformations, at least those of small dimension such as 
shape fluctuations, are dominated by the curvature elasticity and are
essentially ``shear-free''. The same conclusion was obtained from
the observations on both artificial lipid vesicles [11] and on 
erythrocytes [12]. Another argument is given by the measurements 
of the bending modulus ($k'_{11}$). It was reported that the bending 
modulus of lipid membranes is very close, in magnitude, to that 
of human erythrocyte membranes: for phospholipid vesicles 
$k'_{11}=1 \sim 4 \times 10^{-12} \ erg$ [5,13,14]\footnote\dag
{The data cited here are for egg lecithin.
For pure phospholipid membranes, it was found that 
$k'_{11}=4 \sim 5 \times 10^{-13} \ erg$ [17,18]. Adding cholesterol to 
the pure lipid membranes may elevate the magnitude to one order 
higher. For example, the modulus of dimyristoylphosphatidycholine 
containing $50 \ mol\% $ cholesterol has the value 
$k'_{11}=4 \times 10^{-12} \ erg $ [19]. This composition equals 
to that of erythrocyte membranes which contains $50 \ mol\% $ 
phospholipids and $50 \ mol\% $ cholesterol.}
and for red blood cells $k'_{11}=1 \sim 7 \times 
10^{-12} \ erg$ [12,15,16]\footnote\ddag  
{The data cited here correspond to the exciting wave length comparable 
to the cell dimension ($10 \ \mu m$). It was also reported that to 
short wave length excitation ($0.1 \sim 0.3 \ \mu m$), the bending modulus 
of red cell membranes is $1 \sim 2 \times 10^{-13} \ erg$ [20,21]. For a 
discussion about the discrepancy refer to [12].}.

We expect that the theory presented in this paper will be extensively
used in the study of dynamic processes of lipid vesicles, as well as 
biological cells, suspended in aqueous solutions. 
As a special application, one gives in section 3 the hydrostatic 
equations of the membrane in the field of fluid pressure  
and suggests to interpret the complex cell shapes by means of the
analysis of stability, since biological systems are in general 
far from thermodynamic equilibrium. 

\bigskip
\leftline{\bf 2. Connection with liquid surroundings}
\medskip

The bilayer lipid membrane is considered to be adjacent to a
stokesian fluid on each side. The thickness of the membrane is neglected and
thus the ``middle surface" of the membrane coincides with the ``dividing
surface" of the two fluids (see Fig. 1). The position of the two superposed 
geometric surfaces in an Euclidean space is given by the vector 
$$ {\rm r}^{\rm i} = {\rm r}^{\rm i} \left( \theta^1,\,\theta^2,\,t \right) 
\eqno (1) $$ 
where $ \theta^\alpha $ are curvilinear coordinates on the surface 
and $ t $ is the time. Here and throughout the paper, block letters stand 
for vectors or tensors expressed in the inertial coordinates fixed in the 
Euclidean space and italic letters for those in the local bases 
(see reference [1]); Latin suffixes range over $ 1,\,2,\,3 $, while 
the Greek ones over $ 1,\,2 $. 

Denoting the displacement of the surface by $ \zeta $, one defines the 
speed of the displacement movement by
$$ w = {\partial\zeta\over\partial t} \eqno (2) $$ 
Both $\zeta$ and $w$ are measured in the direction normal to the surface
(see the definition of the displacement movement [1]).

\medskip
\leftline{\sl 2.1 Mass balance}
\smallskip

In the absence of exchange with the surroundings, the membrane 
has a constant total mass. The mass conservation principle for an 
arbitrary area $ S $ is written  
$$ {d\over dt} \int\limits_S\!\!\!\int \gamma \, ds = 0 \eqno (3) $$ 
where $ \gamma $ is the surface density of mass.

When the membrane is permeable to certain solutes but not to the  
solvent, the total mass balance is given by  
$$ {d\over dt} \int\limits_S\!\!\!\int \gamma \, ds = - \sum\limits_{k}
\int\limits_S\!\!\!\int \left[ j_{k(n)} \right] \, ds  \eqno (4) $$ 
where $ {\bf j}_k $ is the current density of the $ kth $ considered 
species crossing the membrane. The subscript $ (n) $ denotes the 
normal component of vectors. The square brackets stand for the
discontinuities. For an arbitrary physical quantity $ \phi $ the jump 
at the surface is defined as
$$ \left[ \phi \right] = \phi_{conv} - \phi_{conc} $$
with the subscript $ conc $ for the concave side and $ conv $ for 
the convex side of the surface. 

When osmosis of the adjacent solutions occur, the instantaneous mass 
change is additionally furnished by the absorption flux over the area. 
So one has 
$$ {d\over dt} \int\limits_S\!\!\!\int \gamma \, ds = 
- \sum\limits_{k} \int\limits_S\!\!\!\int \left[ j_{k(n)} \right] \, ds
- \int\limits_S\!\!\!\int 
\left[ \rho \left( u_{(n)} - w \right) \right] \, ds \eqno (5) $$ 
where $ \rho $ is the mass density and $ \bf u $ the velocity of the 
bulk fluids. 

However, the Reynolds transport theorem for two-dimensional fluids  
asserts that [22] 
$$ {d\over dt} \int\limits_S\!\!\!\int \phi \, ds = \int\limits_S\!\!\!\int 
\left[ \dot\phi + \phi \left( {\dot a \over 2a} + v^\alpha_{\;,\,\alpha} 
\right) \right] \, ds \eqno (6) $$ 
where, $ \phi $ is any scalar function of position and time, $ a $ the 
determinant of the metric tensor of the curvilinear coordinate system 
and $ \bf v $ the velocity of the fluids\footnote\dag 
{In this paper, $\bf u$ denotes the velocity of the surrounding fluids, 
whereas $\bf v$ denotes that of the membrane. In the local bases, the
velocity of the membrane (without thickness) is expressed as 
$$ {\bf v} = v^\alpha {\bf e}_\alpha + w {\bf e}_3 $$
with $ {\bf e_3} = {\bf n} $, $\bf n$ being the normal unit vector.}.
The top dot stands for the 
convected derivative and the comma for the covariant derivative 
along the surface. Summation convention is followed here and 
throughout the paper. Besides, the dilation of the area element is 
related to the displacement movement by [9]
$$ {\dot a \over 2a} = - 2Hw \eqno (7) $$ 
$ H $ being the mean curvature of the surface. 
So Eqs.(3-5) may be written in the differential form
$$ \eqalignno{ 
\dot\gamma + \gamma \left( v^\alpha_{\;,\,\alpha} - 2Hw \right) 
& = 0 \cr 
& {\rm (for \ closed \ membrane \ system)} & (8) \cr 
\noalign{\vskip3pt}
\dot\gamma + \gamma \left( v^\alpha_{\;,\,\alpha} - 2Hw \right) 
& = - \sum \limits_{k} \left[ j_{k(n)} \right] \cr
& {\rm (for \ solute \ permeation \ process)} & (9) \cr
\noalign{\vskip3pt}
\dot\gamma + \gamma \left( v^\alpha_{\;,\,\alpha} - 2Hw \right) 
& = - \sum \limits_{k} \left[ j_{k(n)} \right]
- \left[ \rho \left( u_{(n)} - w \right) \right] \cr 
& {\rm (for \ solution \ osmosis \ process)} & (10) \cr} $$
Here we have omitted the lateral diffusion of the particles moving
across the membrane.

\medskip
\leftline{\sl 2.2 General form of balance equations}
\smallskip

Suppose first that no exchange of the physical quantity $ \phi $ occurs 
between the membrane and the bulk phases. Let $ \phi = \gamma \bar\phi^s $. 
One may write then the instantaneous balance in the form
$$ {d\over dt} \int\limits_S\!\!\!\int \gamma {\bar\phi}^s \, ds 
= \int\limits_S\!\!\!\int Q \, ds - \oint\limits_L J^\alpha \, dl_\alpha 
\eqno (11) $$ 
where $ L $ is the periphery of the area $ S $, $ Q $ the production 
of $ \phi $ per unit time per unit area and $ \bf J $ the current 
density of $ \phi $ along the surface. Making use of theorem (6), as 
well as the Stokes theorem, and taking account of (7), one obtains 
$$ \gamma \dot {\bar\phi^s} + \bar\phi^s \left[ \dot\gamma + \gamma 
\left( v^\alpha_{\;,\,\alpha} - 2 Hw \right) \right] = 
Q - J^\alpha_{\;,\,\alpha}  \eqno (12) $$ 

Imagine next that the two bulk phases are in direct contact
with each other without the presence of the membrane between them. The 
condition of compatibility at the dividing surface is given as 
(Kotchine's theorem [9]) 
$$ \left[ \rho \bar\phi^v \left( u_{(n)} - w \right) \right] + 
\left[ j_{(n)} \right] = 0 \eqno (13) $$
where $ \rho \bar\phi^v = \phi $ and $ \bf j $ is the influx 
of $ \phi $ from the volumes to the surface.

Then we consider the membrane being in contact with the surroundings. 
Evidently, in this case the accumulation brought by the flux from 
the volumes makes a part of the total amount of $ \phi $ pertinent 
to the membrane. As the middle surface of the membrane and the 
dividing surface of the two bulks are coincident, one obtains 
naturally the desired balance equation by summing Eqs.(12) and (13)  
$$ \eqalignno{
& \gamma \dot {\bar\phi^s} + \bar\phi^s \left[ \dot\gamma + \gamma \left(
v^\alpha_{\;,\,\alpha} - 2 Hw \right) \right] \cr 
& \hskip .5cm = Q - J^\alpha_{\;,\,\alpha} - \left[ j_{(n)} \right] 
- \left[ \rho \bar\phi^v \left( u_{(n)} - w \right) \right] 
& (14) \cr} $$
In particular, put $ {\bar\phi}^s = {\bar\phi}^v = 1 $, then  
Eq.(14) will be reduced to Eq.(10), if the source of 
substances is null and the lateral diffusion of components is negligible.  

Inserting (8), (9) and (10) respectively into Eq.(14) gives 
the balance equations of $ \phi $, for the closed membrane system
$$ \eqalignno{  \gamma \dot {\bar\phi^s} = 
& Q - J^\alpha_{\;,\,\alpha} - \left[ j_{(n)} \right] 
- \left[ \rho \bar\phi^v \left( u_{(n)} - w \right) \right] & (15) \cr 
\noalign{\noindent\rm for\ the\ solute\ permeation\ process}
\gamma \dot {\bar\phi^s} = & Q - J^\alpha_{\;,\,\alpha} - 
\left[ j_{(n)} \right] 
- \left[ \rho \bar\phi^v \left( u_{(n)} - w \right) \right]
+ \bar\phi^s \sum\limits_k 
\left[ j_{k(n)} \right] & (16) \cr 
\noalign{\noindent\rm and\ for\ the\ solution\ osmosis\ process}
\gamma \dot {\bar\phi^s} = & Q - J^\alpha_{\;,\,\alpha} - 
\left[ j_{(n)} \right] 
- \left[ \rho \bar\phi^v \left( u_{(n)} - w \right) \right] + \cr
& + \bar\phi^s \sum\limits_k \left[ j_{k(n)} \right] +  
\bar\phi^s \left[ \rho \left( u_{(n)} - w \right) \right] & (17) \cr} $$ 

\smallskip

In the following, one considers only the closed membrane system. 

\medskip
\leftline{\sl 2.3 Momentum exchange}
\smallskip

With regard to the momentum exchange, the physical quantity $ \bar\phi $ is 
interpreted as the velocities 
$$ \eqalignno{ 
& \bar\phi^s = v^\alpha t^{\rm i}_\alpha + w {\rm n}^{\rm i} \cr  
& \bar\phi^v = {\rm u}^{\rm i} & (18) \cr} $$ 
where $ \bf n $ is the unit normal vector to the surface 
and $ \bf t $ is the $ 3 \times 2 $ hybrid tensor having the components
$$ t^{\rm i}_\alpha = {\partial {\rm x}^{\rm i} 
\over \partial \theta^\alpha} \eqno (19) $$ 
$ {\rm x}^{\rm i} $ being the inertial coordinates.
The net influx $ \left[ j_{(n)} \right] $ is interpreted as the 
resultant of the forces that the bulk phases exert on the surface
$$ \left[ j_{(n)} \right] = \left[ \varrho \right] \hat {\rm g}_{(n)} 
{\rm n}^{\rm i} - [ p ] {\rm n}^{\rm i} + 
2 \left[ \mu_v {\rm e}^{\rm ij} \right] {\rm n}_{\rm j} \eqno (20) $$  
where $ \varrho $ equals to the mass density $ \rho $ multiplied by
a unit length, $ \hat {\bf g} $ is the gravitational acceleration, $ p $ 
the thermodynamic pressure, $ {\rm e}^{\rm ij} $ the 
rate of strain tensor and $ \mu_v $ the viscosity  
coefficient of the bulk fluids. The source of $ \phi $ is interpreted 
as the gravitational attraction to the membrane
$$ Q = \gamma \hat {\rm g}^{\rm i} \eqno (21) $$
The surface flux $ \bf J $ is interpreted as the surface force inherent 
in the membrane 
$$ J^\alpha = - T^{\alpha\beta} t^{\rm i}_\beta - T^{\alpha 3} 
{\rm n}^{\rm i} \eqno (22) $$ 
where $ \bf T $ is the stress tensor of the membrane. 
With these interpretations, Eq.(15) reads 
$$ \eqalignno{ & \gamma {d\over dt} \left( v^\alpha  t^{\rm i}_\alpha
+ w {\rm n}^{\rm i} \right) \cr  
& \hskip .5cm = \gamma \hat {\rm g}^{\rm i} + \left( T^{\alpha\beta} 
t^{\rm i}_\beta + T^{\alpha 3} {\rm n}^{\rm i} \right)_{,\,\alpha} - \cr 
& \hskip 1.cm - \left[ \varrho \right] \hat {\rm g}_{\rm j} 
{\rm n}^{\rm j} {\rm n}^{\rm i} + [ p ] {\rm n}^{\rm i} - 
2 \left[ \mu_v {\rm e}^{\rm ij} \right] {\rm n}_{\rm j} & (23) \cr} $$ 
In writing (23) we have supposed the continuity of velocity at the surface
$$ {\rm u}_{(n)} = w \eqno (24) $$
Making use of the kinematic formulae [23,24] 
$$ \eqalignno{ \dot t^{\rm i}_\alpha = & \left( w {\rm n}^{\rm i} \right)
_{,\,\alpha} + t^{\rm i}_{\alpha,\,\beta} v^\beta & (25) \cr 
\noalign{\noindent\rm and}
\dot {\rm n}^{\rm i} = & \partial_t {\rm n}^{\rm i} + {\rm n}^{\rm i}
_{\;,\,\beta} v^\beta + w \left\{ \matrix{\rm i \cr \rm j \; k \cr} \right\}
{\rm n}^{\rm j} {\rm n}^{\rm k} \cr 
= & - a^{\alpha\beta} t^{\rm i}_\alpha \left( w {\rm n}^{\rm j} \right)
_{,\,\beta} {\rm n}_{\rm j} + {\rm n}^{\rm i}_{\;,\,\beta} v^\beta \cr 
= & - t^{\rm i}_\alpha \left( a^{\alpha\beta} \partial_\beta w + 
b^\alpha_\beta v^\beta \right) & (26) \cr} $$ 
where $ a_{\alpha\beta} $ and $ b_{\alpha\beta} $ are respectively the
first and the second magnitude of the surface, 
$ \left\{ \matrix{\rm i \cr \rm j \; k \cr} \right\} $ is the second 
kind of Christoffel symbol, one rewrites the first term on the left 
side of (23) as 
$$ \eqalignno{ & \gamma {d\over dt} \left( v^\alpha t^{\rm i}_\alpha + w
{\rm n}^{\rm i} \right) \cr  
& \hskip .5cm = \gamma \left( \dot v^\alpha - 2b^\alpha_\beta v^\beta w -
a^{\alpha\beta} w \partial_\beta w \right) t^{\rm i}_\alpha + \cr 
& \hskip 1cm + \gamma \left( \dot w + v^\alpha \partial_\alpha w +
b_{\alpha\beta} v^\alpha v^\beta \right) {\rm n}^{\rm i} & (27) \cr} $$ 
Additionally, one expands the second term on the right 
side of (23) in the form (see Appendix)
$$ \eqalignno{ 
& \left( T^{\alpha\beta} t^{\rm i}_\beta + T^{\alpha 3} {\rm n}
^{\rm i} \right)_{,\,\alpha} \cr 
& \hskip .5cm = \left( T^{\alpha\beta}
_{\;,\,\alpha(\tau)} - b^\beta_\alpha  T^{\alpha 3} \right) t^{\rm i}_\beta +  
\left( T^{\alpha 3}_{\;,\,\alpha(\tau)} + b_{\alpha\beta} T^{\alpha\beta} 
\right) {\rm n}^{\rm i} & (28) \cr} $$ 
with 
$$ \eqalignno{ T^{\alpha\beta}_{\;,\,\alpha(\tau)} = & \partial_\alpha
T^{\alpha\beta} + {1\over 2a} T^{\alpha\beta} \partial_\alpha a
+ \left\{ \matrix{\beta \cr \sigma \; \alpha} \right\}
T^{\alpha\sigma} \cr 
T^{\alpha 3}_{\;,\,\alpha(\tau)} = 
& \partial_\alpha T^{\alpha 3} + {1\over 2a} T^{\alpha 3}
\partial_\alpha a & (29) \cr} $$ 
where the subscript $ (\tau) $ is used to denote the tangential component 
of vectors or tensors. Then, by using (27) and (28), one splits Eq.(23) 
into the tangential part 
$$ \eqalignno{ & \gamma \left( \dot v^\alpha - 2b^\alpha_\beta v^\beta w -
a^{\alpha\beta} w \partial_\beta w \right) \cr 
& \hskip .3cm = \gamma \hat {\rm g}_{\rm j} t^{\rm j}_\beta a^{\beta\alpha} 
+ T^{\beta\alpha}_{\;,\,\beta(\tau)} - b^\alpha_\beta T^{\beta 3}  
- 2 \left[ \mu_v {\rm e}_{\rm ij} \right] {\rm n}^{\rm j} 
t^{\rm i}_\beta a^{\beta\alpha} & (30) \cr
\noalign{\noindent\rm and the\ normal\ part}
& \gamma \left( \dot w + v^\alpha \partial_\alpha w + 
b_{\alpha\beta} v^\alpha v^\beta \right) \cr
& \hskip .3cm = \gamma \hat {\rm g}_{\rm j} {\rm n}^{\rm j} + 
T^{\beta 3}_{\;,\,\beta(\tau)} 
+ b_{\beta\alpha} T^{\alpha\beta} - \left[ \varrho \right] 
\hat {\rm g}_{\rm j} {\rm n}^{\rm j} + [p] - 
2 \left[ \mu_v {\rm e}_{\rm ij} \right] {\rm n}^{\rm i} {\rm n}^{\rm j} 
& (31) \cr} $$
The rate of strain tensor for the stokesian fluids is given, 
as well known, by
$$ {\rm e}_{\rm ij} = {1\over 2} \left( {\rm u}_{\rm i,\,j} + {\rm u}_{\rm
j,\,i} \right) \eqno (32) $$                              
The stress tensor of the membranes without any chiral structure has the 
components [1]
$$ \eqalignno{ T^{\alpha\beta} =  & - \sigma a^{\alpha\beta} + \left( k'_1 - 
2Hk'_{11} \right) b^{\alpha\beta} + \cr
& + \left( k'_{22} + k'_{24} \right) K a^{\alpha\beta} - \left( \mu - \eta 
\right) 2Hwa^{\alpha\beta} - \cr
& - \eta \left\{ 2w b^{\alpha\beta} - {1\over 2} E^{\alpha\beta\lambda\mu} 
\left(  v_{\lambda,\,\mu} + v_{\mu,\,\lambda} \right) \right\} \cr 
T^{\alpha 3} = & - {\alpha_3 \over 2} b^\alpha_\beta v^\beta & (33) \cr} $$ 
where, the $ k'_i $'s and the $ k'_{ij} $'s are the elastic moduli, 
$ \mu $ and $ \eta $ the dilatation viscosity and the shear viscosity, 
respectively, $ \sigma $ is the surface tension, $ \alpha_3 $ the 
viscosity relevant to the intrinsic rotation of the molecules, $ K $ the 
gaussian curvature, $ \varepsilon^{\beta\alpha} $ the two-parametric 
permutation tensor having the components 
$ \varepsilon^{11} = \varepsilon^{22} = 0,\ \varepsilon^{12} = 
- \varepsilon^{21} = 1/\sqrt a $, $ E^{\beta\alpha\lambda\mu} $ 
the fourth-order transversely isotropic tensor defined as 
$ E^{\beta\alpha\lambda\mu} = a^{\alpha\lambda} a^{\beta\mu} + a^{\alpha\mu} 
a^{\beta\lambda} - a^{\alpha\beta} a^{\lambda\mu} $.
Inserting relations (32) and (33) into Eqs.(30) and (31), one obtains 
finally the explicit form of the equation of motion for the internal flow 
within the membrane 
$$ \eqalignno{ & \gamma \left( \dot v^\alpha - 2b^\alpha_\beta v^\beta w -
a^{\alpha\beta} w \partial_\beta w \right) \cr 
& \hskip .3cm = \gamma \hat {\rm g}_{\rm j} t^{\rm j}_\beta a^{\beta\alpha}
- a^{\beta\alpha} \partial_\beta \sigma + k'_1 a^{\beta\alpha} 
\partial_\beta \left( 2H \right) - \cr 
& \hskip .8cm - k'_{11} \left( b^{\beta\alpha} + 2H a^{\beta\alpha} \right)
\partial_\beta \left( 2H \right) + \left( k'_{22} + k'_{24} \right) 
a^{\beta\alpha} \partial_\beta K - \cr 
& \hskip .8cm - \left( \mu + \eta \right) a^{\beta\alpha} \partial_\beta 
\left( 2H \right)w - 
\left\{ \left( \mu - \eta \right) 2H a^{\alpha\beta} 
+ 2 \eta b^{\alpha\beta} \right\} \partial_\beta w + \cr
& \hskip .8cm + {\eta\over 2} E^{\beta\alpha\lambda\mu} 
\left( v_{\lambda,\,\mu\beta} + v_{\mu,\,\lambda\beta} \right)  
+ {\alpha_3\over 2} \left( 2H b^\alpha_\beta v^\beta - 
K v^\alpha \right) - \cr
& \hskip .8cm - \left[ \mu_v \left( {\rm u}_{\rm i,\,j} + 
{\rm u}_{\rm j,\,i} \right)  \right] {\rm n}^{\rm i} t^{\rm j}_\beta 
a^{\beta\alpha} & (34) \cr 
\noalign{\noindent\rm and\ that\ for\ the\ displacement\ movement\ 
of\ the\ membrane} 
& \gamma \left( \dot w + v^\alpha \partial_\alpha w +
b_{\alpha\beta} v^\alpha v^\beta \right) \cr 
& \hskip .3cm = \gamma \hat {\rm g}_{\rm j}{\rm n}^{\rm j} - 2H \sigma + 
\left( k'_1 - 2H k'_{11} \right) \left( 4H^2 - 2K \right) + 
\left( k'_{22} + k'_{24} \right) 2HK - \cr 
& \hskip .8cm - 4 \left\{ \left( \mu + \eta \right) H^2 - \eta K \right\} w + 
\eta \left( b^{\lambda\mu} - H a^{\lambda\mu} \right) 
\left( v_{\lambda,\,\mu} + v_{\mu,\,\lambda} \right) - \cr 
& \hskip .8cm - {\alpha_3\over 2} 
\left\{ b^\beta_\sigma v^\sigma_{\;,\,\beta} 
+ v^\beta \partial_\beta \left( 2H \right)\right\} - \left[ \varrho \right] 
\hat {\rm g}_{\rm j} {\rm n}^{\rm j} +[ p ] - \cr 
& \hskip .8cm -  \left[ \mu_v \left( {\rm u}_{\rm i,\,j} + 
{\rm u}_{\rm j,\,i} \right) \right] {\rm n}^{\rm i} {\rm n}^{\rm j} 
& (35) \cr} $$

\medskip
\leftline{\sl 2.4 Energy exchange}
\smallskip

With regard to the energy exchange, one might interpret the physical 
quantity $ \bar\phi $ in (15) as the energy of unit mass, the 
surface flux $ \bf J $ as the sum of the current density of heat 
and work, and so on, to get the balance equation. But here we use 
the results obtained before to make the discussion brief.

One has already the energy balance equation of an isolated membrane [1] 
$$ \gamma C_l \dot\Theta = Q^h - {J^h}^\alpha_{\;,\,\alpha} +
T^{\beta\alpha} v_{\alpha,\,\beta} + T^{\beta 3} \partial_\beta w 
\eqno (36) $$ 
with the constitutive relation of heat flux 
$$ {\bf J}^h = - \kappa_\bot \partial_{\rm j} \Theta t^{\rm j}_\alpha
{\bf e}^\alpha - \kappa_\| \partial_{\rm j} \Theta {\rm n}^{\rm j} {\bf e}^3
\eqno (37) $$ 
where, $ C_l $ is the heat capacity per unit mass of the membrane medium in
the static state, $ \Theta $ the temperature, $ Q^h $ the heat source, 
$ \kappa $ the heat conductivity with the subscripts $ \bot $ or $ \|
$ indicating the conduction direction perpendicular or parallel to the normal
vector and $ {\bf e}^k $ are the reciprocal local bases. While the membrane 
is in contact with the bulk fluids, energy current flows from the 
volume phases to the surface in the form of heat and work
$$ - \left[ j_{(n)} \right] = \left[ \kappa_v \Theta_{,\,\rm k} \right]
{\rm n}^{\rm k} - w \left\{ \left[ \varrho \right] \hat {\rm g}_{(n)} - 
[p] + 2 \left[ \mu_v {\rm e}_{\rm ij} \right] {\rm n}^{\rm i}
{\rm n}^{\rm j} \right\} \eqno (38) $$
where $ \kappa_v $ is the heat conductivity of the volume phases. By 
summing (36) and (38), one obtains the energy balance equation
$$ \eqalignno{ \gamma C_l \dot \Theta =  & Q^h + \kappa_\bot \left( 
\partial_{\rm j} \Theta \right)_{,\,\rm k}  \left( {\rm g}^{\rm kj} - 
{\rm n}^{\rm k} {\rm n}^{\rm j} \right) -  \kappa_a 2H 
\left( {\rm n}^{\rm j} \partial_{\rm j} \Theta \right) + \cr 
& + \biggl( - \sigma a^{\alpha\beta} + 
\left( k'_1 - 2Hk'_{11} \right) b^{\alpha\beta} + \biggr. \cr
& \hskip .6cm \biggl. + \left( k'_{22} + k'_{24} \right) K a^{\alpha\beta} 
- \left( \mu - \eta \right) 2Ha^{\alpha\beta} w - \biggr. \cr
& \hskip .6cm \biggl. - \eta \left\{ 2w b^{\alpha\beta} - {1\over 2}
E^{\alpha\beta\lambda\mu} \left( v_{\lambda,\,\mu} +
v_{\mu,\,\lambda}  \right) \right\} \biggr) v_{\alpha,\,\beta} - \cr
& - {\alpha_3 \over 2} b^\beta_\alpha v^\alpha \partial_\beta w +
\left[ \kappa_v \partial_{\rm k}\Theta \right] {\rm n}^{\rm k} - \cr
& - w \biggl\{ \left[ \varrho \right] \hat {\rm g}_{\rm j} {\rm n}^{\rm j} 
- [p] + \left[ \mu_v \left( {\rm u}_{\rm i,\,j} + 
{\rm u}_{\rm j,\,i} \right) \right] {\rm n}^{\rm i} 
{\rm n}^{\rm j} \biggr\} & (39) \cr} $$ 
where $ \kappa_a = \kappa_\| - \kappa_\bot $.
In writing (39), one took relations (33) and (38) into account.

\bigskip
\leftline{\bf 3. Hydrostatic equilibrium}
\medskip

Let $ v^\alpha = w = {\rm u}^{\rm i} = 0 $, Eqs.(34) and (35) 
give
$$ \eqalignno{ & \gamma \hat {\bf g} \cdot {\bf t} 
- {\bf\nabla}_s \sigma + k'_1 {\bf\nabla}_s 
\left(2H \right) + \left( k'_{22} + k'_{24} \right) 
{\bf\nabla}_s K + \cr 
& \hskip .3cm + k'_{11} \left\{ \left( {\bf\nabla}_s {\bf n} 
\right) \cdot {\bf\nabla}_s \left( 2H \right) - 2H 
{\bf\nabla}_s \left( 2H \right) \right\} = 0 & (40) \cr 
\noalign{\noindent\rm and}
& \gamma \hat{\rm g}_{(n)} - 2H \sigma 
+ k'_1 \left( 4H^2 - 2 K \right) 
+ \left( k'_{22} + k'_{24} \right) 2H K - \cr
& \hskip .3cm - k'_{11} 2H \left( 4H^2 - 2K \right) - 
[\varrho ] \hat {\rm g}_{(n)} + [p] = 0 & (41) \cr} $$ 
where the differential operator is defined by
$$ {\bf\nabla}_s = {\bf e}^\beta {\partial\over\partial 
\theta^\beta} \eqno (42) $$

For a non-spherical vesicle, the pressure difference between the interior
and the exterior may be estimated to be of the order of $ 1 \sim 10 
\ dyn \cdot cm^{-2} $ [25], whereas the gravitational attraction 
$ \gamma \hat {\bf g} $ is approximately 
$ 5 \times 10^{-4} \ dyn \cdot cm^{-2} $. So the later
is negligible as an external force. On the other hand, theoretical 
analyses and experiments show that the saddle-shape conformation, 
given by the term of $ k'_{22} + k'_{24} $, can not be observed in
lipid-water systems except when the water content is extremely
low [26]. Therefore Eqs.(40) and (41) can be simplified to 
$$ {\bf\nabla}_s \sigma - k'_1 {\bf\nabla}_s \left( 2H \right)
- k'_{11} \left\{ \left( {\bf\nabla}_s {\bf n} \right) \cdot
{\bf\nabla}_s \left( 2H \right) - 2H {\bf\nabla}_s
\left( 2H \right) \right\} = 0 \eqno (40') $$
and
$$ 2H \sigma - k'_1 \left( 4H^2 - 2K \right) 
+ k'_{11} 2H \left( 4H^2 - 2K \right) +  
[\varrho ] \hat {\rm g}_{(n)} - [p] =0 \eqno (41') $$ 

Equations (40') and (41') serve as the conditions of mechanical equilibrium 
for the membrane-fluid system in the directions tangential and normal to 
the surface, respectively. Eq.(41') gives explicitly the dependence 
of the elasticity of the membrane on the discontinuity of the pressure
field. It is worth while to note that these conditions are different from 
those of thermodynamic equilibrium [25,27,28]. For the latter requires 
further the minimality of the free energy stored in the 
whole elastic surface. When the elasticity vanishes and the buoyancy 
is negligible, Eq.(41') is reduced to the well known Laplace law 
$$ 2H \sigma = [p] \eqno (43) $$

\bigskip
\leftline{\bf 4. Concluding remarks}
\medskip

Eqs.(15), (16) and (17) are given as the general form of 
balance equations for membrane-solution systems. 
Specifically, Eq.(15) is for the process without mass exchange
between the membrane and the liquid surroundings, Eq.(16) for
the process of solute permeation and Eq.(17) for the solvent
osmosis across the membrane.
Particular uses of Eq.(15) yield the equations of motion 
(34) and (35), respectively for the internal molecular flow
inside the membrane and the displacement of the membrane
configuration, as well as the balance equation of energy Eq.(39).
The conditions for the membrane in hydrostatic equilibrium 
with the adjacent fluids are given as a special application
of Eqs.(34) and (35). 
 
\bigskip
\leftline{\bf Acknowledgement}
\medskip

This work is supported by the National Foundation for Natural
Sciences and by the Foundation of the Ministry of Education of China.

\bigskip
\leftline{\bf Appendix. Derivation of Eq.(28)}  
\medskip

Recalling the covariant differentiation [22] 
$$ t^{\rm i}_{\alpha\beta} =
{\partial t^{\rm i}_\alpha\over\partial\theta^\beta}
+ \left\{ \matrix{\rm i\cr j \; k} \right\} 
t^{\rm j}_\alpha t^{\rm k}_\beta
- \left\{ \matrix{\gamma \cr \alpha \; \beta} \right\}
t^{\rm i}_\gamma $$
one writes the first term of (28) as
$$ \eqalignno{
\left( T^{\alpha\beta} t^{\rm i}_\beta \right)_{,\,\alpha} 
= & \partial_\alpha \left( T^{\alpha\beta} t^{\rm i}_\beta \right)
+ \left\{ \matrix{\alpha \cr \alpha \; \mu \cr} \right\} T^{\mu\beta} 
t^{\rm i}_\beta + \left\{ \matrix{\rm i \cr j \; k \cr} \right\} 
t^{\rm j}_\alpha t^{\rm k}_\beta T^{\alpha\beta} \cr
= & t^{\rm i}_\beta \left( \partial_\alpha T^{\alpha\beta} +
\left\{ \matrix{\alpha \cr \alpha \; \mu \cr} \right\} T^{\mu\beta} 
+ \left\{ \matrix{\beta \cr \alpha \; \mu \cr} \right\} T^{\alpha\mu} 
\right) + \cr
& + T^{\alpha\beta} \left( \partial_\alpha t^{\rm i}_\beta +
\left\{ \matrix{\rm i \cr j \; k \cr} \right\} t^{\rm j}_\alpha 
t^{\rm k}_\beta - \left\{ \matrix{ \mu \cr \alpha \; \beta} \right\}
t^{\rm i}_\mu \right) \cr
= & t^{\rm i}_\beta T^{\alpha\beta}_{\;,\,\alpha(\tau)} + 
T^{\alpha\beta} t^{\rm i}_{\beta,\,\alpha} \cr
= & t^{\rm i}_\beta T^{\alpha\beta}_{\;,\,\alpha(\tau)} +
T^{\alpha\beta} b_{\alpha\beta} {\rm n}^{\rm i} \cr
\noalign{\noindent and\ the\ second\ term\ as}
\left( T^{\alpha 3} {\rm n}^{\rm i} \right)_{,\,\alpha}
= & \partial_\alpha \left( T^{\alpha 3} {\rm n}^{\rm i} \right) +
\left\{ \matrix{\rm i \cr j \; k} \right\} t^{\rm k}_\alpha T^{\alpha 3}
{\rm n}^{\rm j} + \left\{ \matrix{\alpha \cr \alpha \mu} \right\}
T^{\mu 3} {\rm n}^{\rm i} \cr
= & {\rm n}^{\rm i} \left( \partial_\alpha T^{\alpha 3} + 
\left\{ \matrix{\alpha \cr \alpha \mu} \right\} T^{\mu 3} \right) +
T^{\alpha 3} \left( \partial_\alpha {\rm n}^{\rm i} + \left\{\matrix{
\rm i \cr j \; k} \right\} t^{\rm k}_\alpha {\rm n}^{\rm j} \right) \cr
= & {\rm n}^{\rm i} T^{\alpha 3}_{\;,\,\alpha(\tau)} + T^{\alpha 3}
{\rm n}^{\rm i}_{\;,\,\alpha} \cr
= & {\rm n}^{\rm i} T^{\alpha 3}_{\;,\,\alpha(\tau)} - T^{\alpha 3}
b^\beta_\alpha t^{\rm i}_\beta \cr} $$
Hence one obtains
$$ \eqalign{
& \left( T^{\alpha\beta} t^{\rm i}_\beta +
T^{\alpha 3} {\rm n}^{\rm i} \right)_{,\,\alpha} \cr
& \hskip .5cm = \left( T^{\alpha\beta}_{\;,\,\alpha(\tau)} - T^{\alpha 3} 
b^\beta_\alpha \right) t^{\rm i}_\beta +
\left( T^{\alpha 3}_{\;,\,\alpha(\tau)} + T^{\alpha\beta} b_{\beta\alpha} 
\right) {\rm n}^{\rm i} \cr} $$

\bigskip
\leftline{\sl References}
\medskip

\item{1.} R. Wang, Colloid. Sur., {\bf 112} (1996) 1.
\item{2.} R. Larter, Chem. Rev., {\bf 90} (1990) 355.
\item{3.} H. C. Pant and B. Rosenberg, Biochim. Biophys. Acta, {\bf 225} 
(1971) 379.
\item{4.} K. Yagisawa, M. Naito, K. I. Gondaira and T. Kambara, 
Biophys. J., {\bf 64} (1993) 1461.
\item{5.} R. M. Servuss, W. Harbich and W. Helfrich, Biochim. Biophys.
Acta, {\bf 436} (1976) 900.
\item{6.} E. Boroske, M. Elwenspoek and W. Helfrich, Biophys. J., {\bf 34} 
(1981) 95.
\item{7.} K. Fricke and E. Sackmann, Biochim. Biophys. Acta, {\bf 803} 
(1984) 145; K. Fricke, K. Wirthensohn, R. Laxhuber and E. Sackmann, 
Eur. Biophys. J., {\bf 14} (1986) 67.
\item{8.} E. Sackmann, H. P. Duwe and H. Engelhardt, Faraday Discuss.
Chem. Soc., {\bf 81} (1986) 281.
\item{9.} C. Truesdell and R. A. Toupin, The Classical Field Theories, 
in S. Flugge (Ed.), Encyclopedia of Physics, Vol.3, No.1, Springer-Verlay, 
Berlin, 1960.
\item{10.} E. A. Evans, Biophys. J., {\bf 13} (1973) 926.
\item{11.} B. Kl\"osgen and W. Helfrich, in R. Lipowsky, D. Richter and 
K. Kremer (Eds.), The Structure and Conformation of Amphiphilic Membranes, 
Springer-Verlag, Berlin, 1992, p. 105.
\item{12.} H. Strey, M. Peterson and E. Sackmann, Biophys. J., {\bf 69} 
(1995) 478.
\item{13.} M. B. Schneider, J. T. Jenkins and W. W. Webb, Biophys. J.,
{\bf 45} (1984) 891; J. Phys. (Paris), {\bf 45} (1984) 1457.
\item{14.} I. Bivas, P. Hanusse, P. Bothorel, J. Lalanne and
O. Aguerre-Chariol, J. Phys. (Paris), {\bf 48} (1987) 855.
\item{15.} M. A. Peterson, Phys. Rev., {\bf 45A} (1992) 4116.
\item{16.} E. A. Evans, Biophys. J., {\bf 43} (1983) 27.
\item{17.} H. Engelhardt, H. P. Duwe and E. Sackmann,
J. Phys. Lett., {\bf 46} (1985) L-395.
\item{18.} J. F. Faucon, M. D. Mitov, P. M\'el\'eard, I. Bivas and
P. Bothorel, J. Phys. (Paris), {\bf 50} (1989) 2389.
\item{19.} H. P. Duwe, J. Kaes and E. Sackmann, J. Phys. (Paris), 
{\bf 51} (1990) 945.
\item{20.} F. Brochard and J. F. Lennon, J. Phys. (Paris), {\bf 36} 
(1975) 1035.
\item{21.} A. Zilker, M. Ziegler and E. Sackmann, Phys. Rev., {\bf 46A} 
(1992) 7998.
\item{22.} R. Aris, Vector, Tensor and the Basic Equations of 
Hydrodynamics, Prentice Hall, Englewood Cliffs, NJ, 1962.
\item{23.} J. C. Slattery, Chem. Eng. Sci., {\bf 19} (1964) 379.
\item{24.} R. Ghez, Sur. Sci. {\bf 4} (1966) 125.
\item{25.} W. Helfrich and H. J. Deuling, J. Phys. (Paris), {\bf 36-C1} 
(1975) C1-327; H. J. Deuling and W. Helfrich, Biophys.J., {\bf 16} 
(1976) 861; J. Phys. (Paris), {\bf 37} (1976) 1335.
\item{26.} A. G. Petrov, M. D. Mitov and A. Derzhanski, Phys. Lett.,
{\bf 65A} (1978) 374.
\item{27.} S. Svetina and B. Zeks, Biomed. Biochim. Acta, {\bf 42}
(1983) 86; Eur. Biophys. J., {\bf 17} (1989) 101.
\item{28.} Zh. C. Ou-Yang and W. Helfrich, Phys. Rev., {\bf 39A} 
(1989) 5280.

\vskip3cm 
\leftline{\sl Figure captions}
\medskip
Fig.1 Illustration of the middle surface of the membrane and the dividing 
surface of the two bulk phases.

\end